\documentclass[preprint,superscriptaddress,amsmath,amssymb]{revtex4}

\usepackage{graphicx}
\usepackage{dcolumn}
\usepackage{bm}

\begin{document}

\title{Accurate characterization of tip-induced potential using electron interferometry}

\author{A. Iordanescu}
\affiliation{IMCN/NAPS, Universit\'e catholique de Louvain (UCLouvain), 1348 Louvain-la-Neuve, Belgium}

\author{S. Toussaint}
\affiliation{IMCN/NAPS, Universit\'e catholique de Louvain (UCLouvain), 1348 Louvain-la-Neuve, Belgium}

\author{G. Bachelier}
\affiliation{CNRS, Institut N\'eel, F-38042 Grenoble, France}

\author{S. Fallahi}
\affiliation{Department of Physics and Astronomy, Purdue University, West Lafayette, IN 47907 USA}
\affiliation{Birck Nanotechnology Center, Purdue University, West Lafayette, IN 47907 USA}

\author{C.G. Gardner}
\affiliation{Birck Nanotechnology Center, Purdue University, West Lafayette, IN 47907 USA}
\affiliation{Microsoft Quantum Purdue, Purdue University, West Lafayette, IN 47907 USA}

\author{M.J.Manfra}
\affiliation{Department of Physics and Astronomy, Purdue University, West Lafayette, IN 47907 USA}
\affiliation{Birck Nanotechnology Center, Purdue University, West Lafayette, IN 47907 USA}
\affiliation{Microsoft Quantum Purdue, Purdue University, West Lafayette, IN 47907 USA}
\affiliation{School of Electrical and Computer Engineering, Purdue University, West Lafayette, IN 47907 USA}
\affiliation{School of Materials Engineering, Purdue University, West Lafayette, IN 47907 USA}

\author{B. Hackens}
\affiliation{IMCN/NAPS, Universit\'e catholique de Louvain (UCLouvain), 1348 Louvain-la-Neuve, Belgium}

\author{B. Brun}
\affiliation{IMCN/NAPS, Universit\'e catholique de Louvain (UCLouvain), 1348 Louvain-la-Neuve, Belgium}

\date{\today}

\begin{abstract}
Using the tip of a scanning probe microscope as a local electrostatic gate gives access to real space information on electrostatics as well as charge transport at the nanoscale, provided that the tip-induced electrostatic potential is well known. Here, we focus on the accurate characterization of the tip potential, in a regime where the tip locally depletes a two-dimensional electron gas (2DEG) hosted in a semiconductor heterostructure. Scanning the tip in the vicinity of a quantum point contact defined in the 2DEG, we observe Fabry-Pérot interference fringes at low temperature in maps of the device conductance. We exploit the evolution of these fringes with the tip voltage to measure the change in depletion radius by electron interferometry. We find that a semi-classical finite-element self-consistent model taking into account the conical shape of the tip reaches a faithful correspondence with the experimental data.
\end{abstract}

\maketitle

Scanning Gate Microscopy (SGM) was invented more than 20 years ago\cite{doi:10.1063/1.117801}, with the objective to probe electron transport at the local scale inside confined electronic systems. SGM consists in locally altering the potential landscape experienced by charge carriers within an electronic device using the biased metallic tip of an Atomic Force Microscope (AFM), while recording the induced changes in the device conductance \cite{sellier2011imaging}.
SGM was first developed to study electronic transport in high mobility two-dimensional electron gases (2DEGs) buried in III-V heterostructures.
In these systems, direct probing of the local electronic density using scanning tunnelling microscopy is prevented by the insulating layer separating the 2DEG from the surface.
The first impressive breakthroughs provided by the SGM technique were the observation of branched electron flow within the 2DEGs \cite{topinka2001coherent}, as well as the ability to image electron wave-functions \cite{topinka2000imaging}. Since then, many groups developed SGM setups, and it proved a very useful tool to investigate mesoscopic transport at the local scale in various systems, such as  quantum dots\cite{fallahi2005imaging,bleszynski2007scanned,huefner2009scanning}, quantum rings\cite{hackens2006imaging,martins2007imaging}, magnetic focusing geometries\cite{aidala2007imaging,bhandari2016imaging}, quantum Hall systems\cite{hackens2010imaging,paradiso2012imaging,connolly2012unraveling,martins2013scanning,martins2013coherent}, and even to explore subtle electron-electron interaction effects\cite{crook2006conductance,jura2010spatially, brun2014wigner,brun2016electron,iagallo2015scanning}. In the past decade, SGM has also been used to provide real-space data on transport through graphene mesoscopic devices\cite{schnez2010imaging,pascher2012scanning,garcia2013scanning,cabosart2017recurrent,dou2018imaging,brun2019imaging}.

In all the above-mentioned cases, SGM relies on measuring the evolution of a device transport property under the influence of an external perturbation, \emph{i.e.} the tip-induced electrostatic potential. The accurate knowledge of this perturbation potential is therefore a crucial issue in the interpretation of the SGM signal. Usually, this potential is estimated in the experiment using its direct effect on transport through a Quantum Point Contact (QPC)\cite{steinacher2015scanning} or a quantum dot\cite{pioda2004spatially,gildemeister2007measurement,liu2015formation}, when the polarized tip scans in the vicinity of the device.
However the latter method suffers from the screening from the top metallic gates, which has been shown to significantly distort the tip-induced potential\cite{schnez2011relevance}.

In the present work, we devise an original way to precisely evaluate the size of the tip-induced depletion region in a high mobility 2DEG, relying on electron interferometry.
We apply a negative voltage to the metallic tip of an AFM to locally deplete the 2DEG, and form a Fabry-P\'erot (FP) cavity between a QPC and the depleted area below the tip.
Following the evolution of a single interference fringe with the tip voltage, we can precisely measure the radius of the depleted region.
This approach is advantageous compared to previous techniques, as the tip-induced perturbation is measured in a pristine area of the 2DEG, without suffering from metallic gate screening. We justify the assumptions underlying the experimental method using tight-binding simulations. Finally, we propose different electrostatic models to reproduce the tip-induced depletion region, in a semi-classical approximation. We demonstrate that the potential induced by a charged sphere and screened by the 2DEG correctly predicts the depletion threshold, but fails to describe the depleted area radius. Finally, we show that modeling the tip as a cone and calculating self-consistently the electrostatic potential reproduces very well the experimental trend.

We study a QPC \cite{vanWees1988,wharam1988} defined using metallic gates deposited on top of an $\rm Al_{0.3}Ga_{0.7}As/GaAs$ heterostructure hosting a 2DEG with a density $n_{s}\rm = 2.53\ \times\ 10^{15}\  m^{-2}$ and mobility $\mu \rm = 3.25\ \times\ 10^{6}\ cm^{2}/(V.s)$, located $d\rm = 57\ nm$ beneath the surface (see Fig.1a). The two metallic top gates are separated from each other by a gap of 300 nm. The sample is thermally anchored on the cold finger of a dilution refrigerator with a base temperature below 100 mK. The QPC conductance $G$ is measured using a four contacts lock-in technique at low frequency (77 Hz). This method consists in polarising the device with an AC voltage $V_{AC}$, typically $\rm 10\ \mu V$, and simultaneously measuring the current $I$ flowing through the device and the voltage drop $V$ across the device. $G=I/V$ is plotted in Fig.1b as a function of the voltage $V_{G}$ applied on the top metallic gates. As $V_{G}$ decreases towards negative values, $G$ exhibits plateaus at integer multiples of $\rm 2e^{2}/h$, corresponding to the number of transverse quantum modes transmitted through the 1D channel between the gates. All subsequent scanning gate measurements were obtained with one quantum mode transmitted through the QPC.

To perform SGM measurements, a metallic tip (a Pt-coated AFM tip provided by $\mu$-masch, model HQ:CSC17/PT\cite{umaschsite}), acting as a local and movable gate, is brought in close proximity to the device surface, at a tip-surface vertical distance $d_{TIP}\rm = 30\ nm$. Applying a negative voltage $V_{TIP}$ on the tip induces a local perturbation for conduction electrons which in turns alters the device conductance. $G$ is recorded as function of $X$ and $Y$ relative tip coordinates, yielding a SGM map. When a sufficiently large negative voltage is applied on the tip, the SGM map reveals a single rather straight branch of reduced conductance aligned with the QPC transport axis, decorated with transverse periodic oscillations. This is illustrated in Fig.1c, showing the SGM map acquired for $V_{TIP}\rm = -6\ V$ on a rectangular region located next to the QPC (the QPC is located 500 nm beyond the left edge of the figure).

\begin{figure} [htbp]
\center
\includegraphics[width=8 cm]{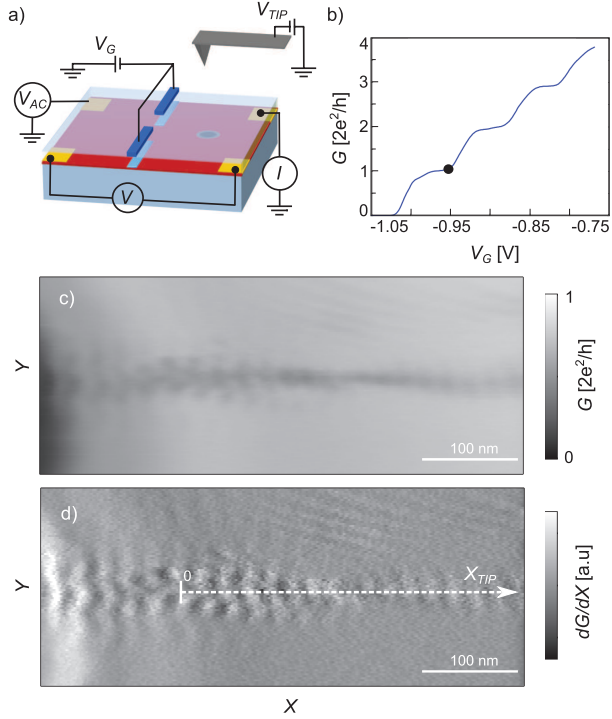}
\caption{(a) Schematic illustration of the scanning gate setup~: top metallic gates defining the QPC are represented in blue, the 2DEG plane in red, and contacts in yellow. (b) Low temperature electrical conductance of the QPC vs $V_{G}$ in units of $\rm 2e^{2}/h$. The black dot indicates the QPC polarization used for the SGM mappings. (c) SGM map of the QPC conductance acquired for $V_{TIP}\rm = -6\ V$ and $d_{TIP}\rm = 30\ nm$, with a voltage $V_{G}\rm = -0.95\ V$ applied on the top metallic gates. (d) Numerical derivative of the data in (c), with respect to $X$. The horizontal axis are matched. The white dashed line corresponds to the $X_{TIP}$ axis in the next figures, with $X_{TIP} = 0$ corresponding to the extremity of the left arrow.}
\end{figure}

The commonly accepted interpretation for the origin of periodic oscillation is based on the formation of a Fabry-P\'erot-like interferometer for electrons \cite{topinka2001coherent,leroy2005imaging}. In this picture, the two mirrors forming the Fabry-P\'erot (FP) cavity are the QPC on one side, and the tip-induced depleted region on the other side. Shifting the tip position successively switches the interference condition between constructive and destructive, leading to an oscillating contrast in the conductance map. The oscillation period should then be given by half of the Fermi wavelength, $\lambda_{F}$. From Fig.1d one can extract $\lambda_{F}/2\rm \simeq 20\ nm$, close to the expected value, given the electronic density ($\lambda_{F}/2 = \sqrt{2\pi/n_{s}}/2 = \rm  24\ nm$), which is consistent with the FP interpretation. 

In the remainder of this paper we will exploit the interference pattern to extract quantitative information on the shape of the tip-induced perturbation. The key data is plotted in Fig.2a, showing the evolution of the interference patterns as a function of the tip voltage while scanning along the white dashed line in Fig.1d, corresponding to the $X_{TIP}$ axis.  The first quantity that can be extracted from this data is the radius of the tip-induced depletion region $R_{TIP}$. In principle, to keep constant the size of the FP cavity when $V_{TIP}$ evolves towards more negative values, the tip has to be moved away from the QPC (as illustrated in Fig.2b). Hence, each iso-conductance line in the interference pattern actually corresponds to an iso-sized FP cavity: as one follows an interference line, a variation of the tip voltage $\Delta V_{TIP}$ beyond the onset of 2DEG depletion translates directly into a variation of $R_{TIP}$, measured as $\Delta X_{TIP}$, the shift in tip position to keep a constant cavity size ($R_{TIP} \rm = 0\ nm$ corresponds to the onset of the interference, see below). For example, when changing $V_{TIP}$ from -4.5 V (onset of depletion, yellow dot in Fig.2a) to -8 V ($\Delta V_{TIP} \rm = -3.5\ V$), the required change in tip position to stay on the same interference fringe is $\Delta X_{TIP} \rm \sim 120\ nm$ from the data in Fig. 2a (consider \emph{e.g.} the yellow and green dots), leading to a tip-induced depletion region with a radius $R_{TIP} \rm \sim 120\ nm$. 

\begin{figure} [htbp]
\center
\includegraphics[width=8 cm]{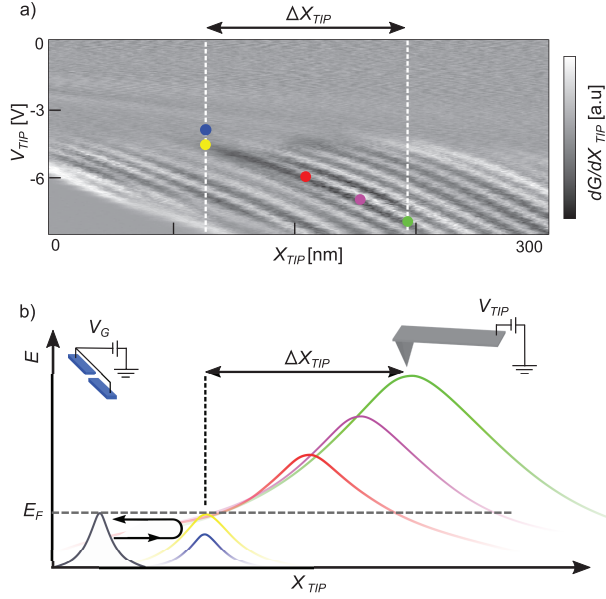}
\caption{(a) Derivative of $G$ with respect to $X_{TIP}$, along the white dashed line in Fig.1d, plotted as a function of $V_{TIP}$, for a tip-sample distance $d_{TIP}\rm = 30\ nm$ and a gate voltage $V_{G}\rm = -0.92\ V$. Except for the blue dot (corresponding to $V_{TIP}\rm = -4\ V$), the other coloured dots correspond to $(X_{TIP},V_{TIP})$ coordinates following the same interference line: yellow dot for $V_{TIP}\rm = -4.5\ V$, red dot for $V_{TIP}\rm = -6\ V$, pink dot for $V_{TIP}\rm = -7\ V$ and green dot for $V_{TIP}\rm = -8\ V$. (b) Schematics of the energy landscape induced by the top gates (dark curve), and by the tip (coloured curves) for the same $(X_{TIP},V_{TIP})$ coordinates indicated in (a). For $V_{TIP}\rm = -4\ V$, the tip induced perturbation is not strong enough to reach the Fermi energy $E_{F} \rm = 4.5\ meV$ and electrons are not backscattered.}
\end{figure}

This experimental method to determine $R_{TIP}$ relies on two main assumptions: (i)  interference fringes are observed as soon as the maximum of the tip-induced perturbation reaches the Fermi energy (\emph{i.e.} the depletion threshold), (ii) the turning point of the electrons at the tip-induced depletion region follows the exact depletion limit. To justify both non-trivial assumptions, we perform tight-binding simulations using the Kwant python package\cite{groth2014kwant} (see supplementary material section I for a detailed description of the method). We scale all the energies and distances to match the experimental conditions. We model the QPC gate potential using the method proposed by Davies et al.\cite{davies1998physics}, and let only one single mode be transmitted through the QPC. We model the tip using an approximate solution for the potential $\phi_{TIP} (r)$ induced at a distance $r$ by a screened charged sphere of radius $R_{S}$, in the Thomas-Fermi approximation\cite{ihn:2010}:
\begin{equation}
\phi_{TIP} (r) = \frac{R_{S} V_{tip}}{\epsilon_r}\int_0^{\infty} q ~ J_0(qr)\frac{e^{-qd}}{q+q_{TF}} ~dq
\end{equation}
where $J_0$ is the zeroth-order Bessel function, and $q_{TF}$ the Thomas-Fermi wave-vector, that we assume to be 2/$a_0$, with $a_0 = \frac{4\pi \epsilon \epsilon_{0} \hbar^2}{m^{*} e^{2}}\simeq 10~ \sf nm$, the effective Bohr radius for electrons in GaAs \cite{liu2015formation,davies1998physics, Krcmar2002,Stern1967}. The electrostatic potential landscape in our simulation is shown in Fig. 3a.

We calculate the total system transmission as a function of the position of this perturbation potential at a variable distance $X_{TIP}$ from the QPC, $X_{TIP} = 0$ nm corresponding to 500 nm away from the QPC (the $X_{TIP}$ axis corresponds to the black dashed line in Fig.3a). We differentiate the calculated transmission versus $X_{TIP}$ and plot the result in Fig.3b, as a function of $X_{TIP}$ and the maximum potential induced by the tip in the 2DEG plane $\phi_{MAX}$, normalized to $E_F$. The result appears very similar to the experiment (Fig.2a), and the interference fringes start to be contrasted below a  voltage threshold $\phi_{MAX}$ very close to the Fermi energy. For $\lvert \phi_{MAX}\lvert > E_F$ the fringe contrast is constant. 

This provides a validation for hypothesis (i): the threshold for the emergence of FP interferences indeed corresponds to the 2DEG depletion. We also plot in Fig.3b the depletion radius $R_{TIP}$ found from Eq.(1) (red curve). Below the depletion threshold and when the depletion spot is well defined and large enough, the FP interference fringes evolve in a way that exactly matches the evolution of the depletion zone, indicating that the turning point is indeed the limit of the depleted area. This simulation therefore also justifies assumption (ii). A discrepancy can however be noted when the tip-induced potential is very close to the Fermi energy, where assumption (ii) does not hold anymore, since the turning point is not given by the depleted limit, either when the tip-induced potential does not deplete the 2DEG but generates efficient backscattering ($\phi_{MAX}\gtrsim E_F$) or when the tip just slightly depletes the 2DEG and the wave-function leaks into the shallow-depleted region ($\phi_{MAX}\lesssim E_F$). This slight discrepancy induces a few percent of uncertainty in the evaluation of the total depletion spot radius but leaves unchanged the conclusions regarding the ability to precisely follow the depletion spot evolution with tip voltage at sufficiently negative $V_{TIP}$.

\begin{figure} [htbp]
\label{fig3}
\center
\includegraphics[width=8 cm]{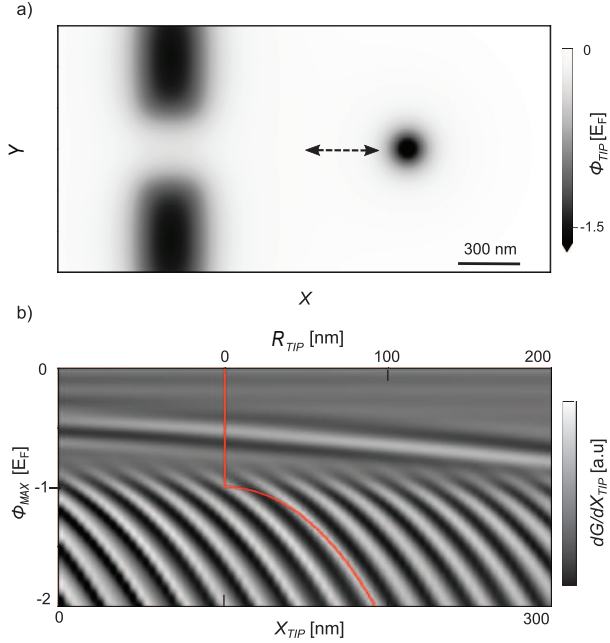}
\caption{(a) Illustration of the electrostatic potential below the finger gates and the tip (dark areas are depleted). (b) Line profiles of the derivative of the simulated $G$ along the black dashed line in (a) (\emph{i.e.} the $X_{TIP}$ axis), as a function of the maximum of the tip-induced perturbation, normalized in units of Fermi energy. The red curve corresponds to the radius of the depletion region ($R_{TIP}$, top axis) calculated with Eq.(1).}
\end{figure}

Next, we compare the outcome of simple electrostatic models of the tip perturbation, taking into account the screening of the 2DEG, with the experimental data. We consider first the analytical tip-2DEG model described by Eq.(1). In this first model (1, in Fig.4a), the conductive sphere is positioned at a vertical distance $d_{TIP - 2DEG}$ from the 2DEG and we neglect the top dielectric layers of $\rm Al_{0.3}Ga_{0.7}As$ and $\rm GaAs$. Fig.4b (black continuous curve) shows the electrostatic potential profile calculated in the 2DEG, for a distance $d_{TIP - 2DEG}\rm = 87\ nm$, a radius $R_{S}\rm = 50\ nm$, while taking $\epsilon_{r}\rm = 10.62 $ and $\ q_{TF} \rm = 0.2\ nm^{-1}$ for the substrate $\rm GaAs$ layer. When compared with the experimental data in Fig. 4c, the shape of the tip - induced potential obtained using Eq.(1) (continuous black curve in Fig.4c) does not correspond to the iso-conductance lines in the FP interference pattern. This can be explained in the light of the approximations used in this model: the dielectric layers above the 2DEG are neglected, as well as the full shape of the tip (assuming that it has a spherical shape). Moreover, changes in the screening of the tip potential due to the emergence of a tip-induced depleted region are not taken into account. In the real device, the 2DEG is housed between a top $\rm Al_{0.3}Ga_{0.7}As$ layer with a thickness $d$ and a $\rm GaAs$ substrate, the experimental AFM tip has a cone-like shape as illustrated in the inset of Fig.4b, and screening phenomena are more complex than the simple description given by Eq.(1). 

\begin{figure} [htbp]
\center
\includegraphics[width=8 cm]{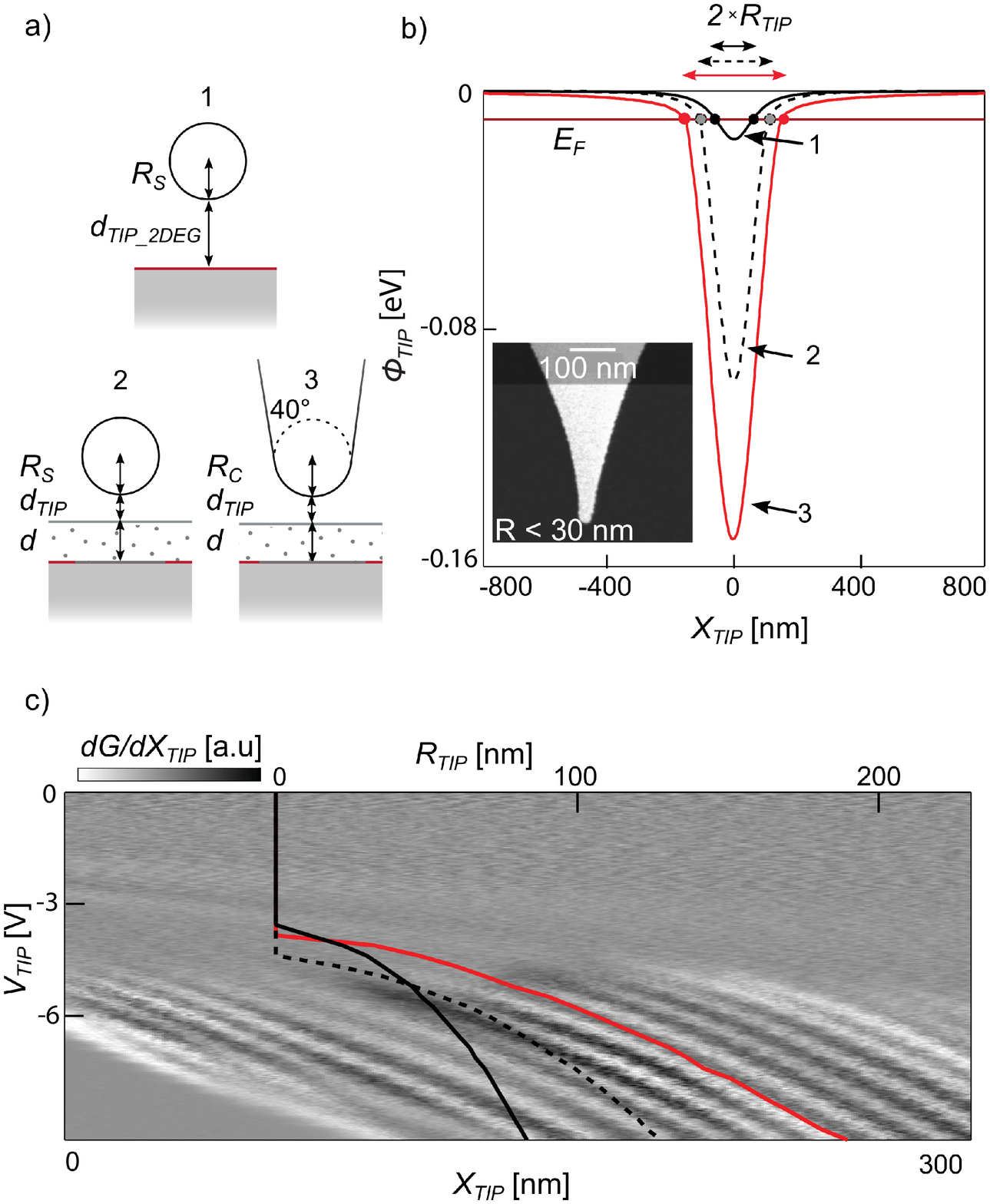}
\caption{(a) Illustration of the three tip - 2DEG models: the red line corresponds to the 2DEG layer, the gray layer corresponds to the $\rm GaAs$ layer while the dotted layer corresponds to the $\rm AlGaAs$ layer. (b) Tip potential profiles traced for the three models presented in (a), for $V_{TIP}\rm = -7\ V$ : (1) a spherical tip in an analytical non-self-consistent framework (continuous black line), (2) a spherical tip in a finite element self-consistent model (dotted black line), and (3) a conical tip in a finite element self-consistent model (red line). The inset shows a side-view electron micrograph of a PtIr tip similar to the one used in the experiment\cite{umaschsite}. The horizontal line corresponds to $E_{F}\rm = -9.6\ meV$. The crossing between the Fermi energy and the tip potential profile determines the diameter tip-induced depletion area. (c) Evolution of line profiles of the derivative of conductance with respect to $X_{TIP}$, recorded as function of $V_{TIP}$ while the tip scans at $d_{TIP}\rm = 30\ nm$. The curves represent the tip-induced depletion radius ($R_{TIP}$, top axis) extracted for the three models presented in (a).}
\end{figure}

To determine the shape of the tip-induced perturbation in a scenario closer to the experimental situation, we simulated the associated electrostatic problem using the Comsol$^\circledR$ software (see supplementary material section III for a detailed description of the method). It consists in finite elements simulation of the electromagnetic field in a region of space where the metallic parts, the dielectric, the doping layer and the 2DEG are defined. Then the Poisson equation is solved by successive iterations which in turn provides the local potential as well as the local electronic density in the depleted 2DEG. In this scenario, the 2DEG lies at $d \rm = 57\ nm$ beneath the surface, between the $\rm Al_{0.3}Ga_{0.7}As$ layer and the $\rm GaAs$ substrate. We assume a relative permittivity $\epsilon_{r}\rm = 10.17 $ for the $\rm Al_{0.3}Ga_{0.7}As$ layer. We neglect the $\rm GaAs$ substrate permittivity ($\epsilon_{r}\rm = 10.62 $), as the layer is considered infinitely thin in the simulations. The doping layer is modeled as a uniformly charged plane, with a density $\rm 2.53\ x\ 10^{15}\ m^{-2}$, insensitive to the local potential, and coinciding with the 2DEG plane. When no voltage is applied neither on gates nor on the tip, the electron density is $\rm 2.53\ x\ 10^{15}\ m^{-2}$.

We consider two models for the tip, as represented in Fig.4a: a sphere with a radius $\ R_{S}$ (model 2) and a cone with a spherical apex with curvature radius $R_{C}\rm = 50\ nm$ and $\rm 40^{\circ}$ full tip cone angle (model 3), both placed at $d_{TIP}\rm = 30\ nm$ from the surface of the sample. For each model, we extract the radius of the tip-induced depletion from the simulated electrostatic potential profile in the 2DEG plane (shown in Fig.4b),and compare it to the outcome of the classical electrostatic model discussed above and to the experimental data (Fig.4c). Compared with model (1), we observe that $R_{TIP}$ vs $V_{TIP}$ estimated using model (2) is closer to the experimental data (Fig.4c, dashed line): it indeed yields a faster evolution of $R_{TIP}$ with $V_{TIP}$, due to the absence of screening of the tip potential by the depleted region below the tip.
However, model (2) still underestimates the depletion spot size compared to the experimental data: the variation of $R_{TIP}$ with $V_{TIP}$ (the dashed line in Fig. 4c) is slower than the evolution of FP interferences.  The model (3) (red curve in Fig.4b-c) yields the most faithful estimate of $R_{TIP}$ with $V_{TIP}$ compared to the experimental result, \emph{e.g.} for a change in tip voltage of $\Delta V_{TIP} = 3.5 \rm\ V$  in the range considered in the experiment (from -4.5 V to -8 V), the increase of $R_{TIP} \rm \sim 120\ nm$, \emph{i.e.} the same value estimated above from the experiment. Furthermore, the non-linear shape of the calculated tip - induced depletion fully reproduces isophase interference profiles observed in the experiment. The last model can therefore serve as reference to evaluate the tip-induced depletion region shape and size.

In conclusion, we performed electron interferometry with a scanning gate microscope, in order to precisely evaluate the size of the depletion radius induced by a polarized SGM tip. We showed that the evolution of interference fringes allows to accurately estimate the depletion radius dependence on tip voltage, and justified
this approach using tight-binding simulations of quantum transport.
A simple electrostatic model of the potential created by a charged sphere and screened by a 2DEG is sufficient to predict the depletion threshold, but
underestimates the size of the depletion spot.
Finally, we showed that a complete modelling of the tip geometry including its conic tail accurately describes the tip-induced depletion region.
This provides new guidelines to choose the best approach to model SGM experiments, in particular when the tip is used as a tunable and movable depleting scatterer. \cite{Kozikov_16,Toussaint_18}.

\begin{acknowledgments}
The present research was partly funded by the F\'ed\'eration Wallonie-Bruxelles through the ARC Grant No. 16/21-077, and by the F.R.S-FNRS through the Grant No. J008019F. B.B. (research assistant), B.H. (research associate) and A.I. (FRIA fellowship)
acknowledge financial support from the F.R.S.-FNRS of Belgium.
\end{acknowledgments}

The data that support the findings of this study are available from the corresponding author upon reasonable request.

\bibliography{references}

\end{document}